\documentclass[prb,letterpaper,twocolumn,showpacs,floatfix,superscriptaddress,amsmath,amsfonts,amssymb,preprintnumbers,citeautoscript]{revtex4}
\pdfoutput=1
\usepackage[T1]{fontenc}\usepackage[latin1]{inputenc}
\usepackage{dcolumn,graphicx,color,hvfloat,booktabs,microtype,afterpage} 
\usepackage[charter]{mathdesign}
\usepackage{sidecap}
\renewcommand{\figurename}{Fig.}
\renewcommand{\tablename}{Table}
\makeatletter\renewcommand{\fnum@figure}[1]{\figurename~\thefigure~(color online).}\makeatother
\makeatletter\renewcommand{\fnum@table}[1]{\tablename~\thetable.}\makeatother

\newcount\hh \newcount\mm
\hh=\time \divide\hh by 60
\mm=\hh \multiply\mm by 60 \mm=-\mm
\advance\mm by \time
\def\now{\number\hh:\ifnum\mm<10{}0\fi\number\mm}

\usepackage[colorlinks,plainpages=false,linkcolor=blue,urlcolor=blue,citecolor=blue,pdfpagemode=UseNone,pdfstartview=FitBH]{hyperref}

\begin{document}\pagestyle{plain}

\title{Quantitative assessment of pinning forces and the superconducting gap in NbN thin films\\from complementary magnetic force microscopy and transport measurements}

\author{T.~Shapoval}\email[Corresponding author: \vspace{4pt}]{t.shapoval@ifw-dresden.de}
\affiliation{IFW Dresden, Institute for Metallic Materials, P.\,O.\,Box 270116, D-01171 Dresden, Germany.}

\author{H.~Stopfel}
\affiliation{IFW Dresden, Institute for Metallic Materials, P.\,O.\,Box 270116, D-01171 Dresden, Germany.}

\author{S.~Haindl}
\affiliation{IFW Dresden, Institute for Metallic Materials, P.\,O.\,Box 270116, D-01171 Dresden, Germany.}

\author{J.\,Engelmann}
\affiliation{IFW Dresden, Institute for Metallic Materials, P.\,O.\,Box 270116, D-01171 Dresden, Germany.}

\author{D.\,S.\,Inosov}
\affiliation{Max-Planck-Institut für Festkörperforschung, Heisenbergstraße 1, D-70569 Stuttgart, Germany}

\author{B.~Holzapfel}
\affiliation{IFW Dresden, Institute for Metallic Materials, P.\,O.\,Box 270116, D-01171 Dresden, Germany.}

\author{V.~Neu}
\affiliation{IFW Dresden, Institute for Metallic Materials, P.\,O.\,Box 270116, D-01171 Dresden, Germany.}

\author{L.~Schultz}
\affiliation{IFW Dresden, Institute for Metallic Materials, P.\,O.\,Box 270116, D-01171 Dresden, Germany.}




\begin{abstract}
\noindent Epitaxial niobium-nitride thin films with a critical temperature of $T_{\rm c}=\!16$\,K and a thickness of 100\,nm were fabricated on MgO\,(100) substrates by pulsed laser deposition. Low-temperature magnetic force microscopy (MFM) images of the supercurrent vortices were measured after field cooling in a magnetic field of 3\,mT at various temperatures. Temperature dependence of the penetration depth has been evaluated by a two-dimensional fitting of the vortex profiles in the monopole-monopole model. Its subsequent fit to a single $s$-wave gap function results in the superconducting gap amplitude $\Delta(0)=(2.9\pm0.4)$\,meV = $(2.1 \pm 0.3)\,k_{\rm B}T_{\rm c}$, in perfect agreement with previous reports. The pinning force has been independently estimated from local depinning of individual vortices by lateral forces exerted by the MFM tip and from transport measurements. A good quantitative agreement between the two techniques shows that for low fields, $B \ll \mu_0H_{\text{c2}}$, MFM is a powerful and reliable technique to probe the local variations of the pinning landscape. We also demonstrate that the monopole model can be successfully applied even for thin films with a thickness comparable to the penetration depth.
\end{abstract}

\maketitle

\vspace{-5pt}\section{Introduction}\vspace{-5pt}

\noindent Vortex pinning is an important characteristic of type~II superconductors that allows tuning its properties on purpose without changing the chemical composition. Hence, the interpretation of pinning mechanisms~\cite{Bla94}, the search for artificial defects with high pinning potentials~\cite{Fol07}, and commensurable pinning effects by ordered arrays of defects~\cite{Mos96,RefA, Ala09, RefC, Hof08, Sha10} remain in the focus of basic research and application-based engineering. On the one hand, many high-power applications require materials with high pinning~\cite{Fol07}, i.e.\,high critical current density. On the other hand, logical applications (i.e.\,fluxonics devices \cite{Nor04}) benefit from low-pinning materials, which can be locally modified by introducing strong pinning sites in order to tune the dynamics of vortices and rectify their motion~\cite{Sil07}.

A common way to investigate the pinning strength is to measure the sample's response to an applied magnetic field or current, using magnetometry or transport measurements, respectively. These global methods probe the average value of the pinning force in a material including the collective dynamics of the elastic vortex lattice. However, the local modulations of the pinning landscape originating from different natural or artificial defects remain inaccessible to these techniques. This challenge can be addressed by local imaging methods, such as low-temperature magnetic force microscopy (LT-MFM), which is capable of correlating the superconducting (SC) vortex positions with the distribution of micro- or nanostructural defects. It effectively combines the non-invasive imaging of flux lines with the ability to manipulate individual vortices by the stray field of the magnetic tip, offering a direct access to the local pinning force~\cite{Str08}. Nevertheless, reconciling the results of local and global measurements often represents a challenge.

In this paper, we will estimate the pinning forces in niobium nitride (NbN) thin films using two complementary methods. Local measurements by LT-MFM will be directly compared to transport measurements. Thin films of NbN have been chosen due to their high critical temperature, $T_{\rm c}\approx16\,\text{--}\,17$\,K, that makes them suitable for LT-MFM studies in a wide temperature range. Moreover, this conventional superconductor has attracted attention due to its recent applications in sensitive SC bolometers \cite{Sem05}.

We will also evaluate the temperature dependence of the magnetic penetration depth, $\lambda(T)$, by performing a two-dimensional fit of the vortex profiles within the monopole-monopole model. Subsequently, we will use these values to estimate the superconducting energy gap of the NbN film, which we will compare to the direct tunneling measurements from the published literature.

\makeatletter\renewcommand{\fnum@figure}[1]{\figurename~\thefigure.}\makeatother
\begin{figure}[b]
\includegraphics[width=\columnwidth]{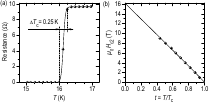}\vspace{-0.3em}
\caption{(a)~Zero-field resistive SC transition with a full width of ${\scriptstyle\Delta}T_{\rm c}\approx0.25$\,K. (b)~Second critical field, $\mu_0 H_{c2}$, as a function of the reduced temperature, $t=T/T_{\rm c}$. Points are the experimental values determined from transport measurements, whereas the solid lines are empirical fits \cite{Tin96}.}\vspace{-1.37em}
\label{Fig:Figure1}
\end{figure}
\makeatletter\renewcommand{\fnum@figure}[1]{\figurename~\thefigure~(color online).}\makeatother

\vspace{-5pt}\section{Experimental details}\vspace{-5pt}

Epitaxial NbN thin films with a thickness $d=100$\,nm were fabricated on single-crystalline MgO\,(100) substrates by pulsed laser deposition (PLD) using a Nb (99.95\,$\%$) target in N$_2$ atmosphere with a pressure of $5\cdot10^{-2}$\,mbar. The base pressure in the chamber was 10$^{-9}$\,mbar. Prior to the film preparation, the Nb deposition rate was measured with an \textit{Inficon XTM/2} rate monitor. We used a KrF excimer laser (\textit{Lambda-Physik}) with a wavelength of 248\,nm and a pulse duration of 25\,ns. The substrates were heated up to 500$^{\circ}$C during deposition. As the on-axis PLD process leads to the formation of droplets on the surface, which pose a severe problem to the MFM scanning tip, a polishing technique \cite{Sha08} was applied to remove these obstacles, providing a peak-to-valley roughness below 5\,nm. X-ray diffraction patterns showed (00$l$) peaks (simple cubic structure) similar to previous reports~\cite{Tre95}. The best samples exhibit a $T_{\rm c}$ (offset) at 16\,K and a sharp resistive SC transition with a width ${\scriptstyle\Delta}T_{\rm c}\approx0.25$\,K [Fig.\,\ref{Fig:Figure1}\,(a)]. The temperature dependence of the second critical field, $\mu_0 H_{\text{c}2}$, determined from transport measurements, is shown in panel (b) of the same figure.

The LT-MFM measurements have been performed using a commercial scanning-probe microscope (\textit{Omicron Cryogenic SFM})~\cite{Sha07}. We have used an MFM cantilever (\textit{Nanoworld MFMR}) that possesses a force constant $k\approx2.8$\,N/m and a resonance frequency $f_0\approx80$\,kHz. For the transport measurements, a 100\,$\mu$m-wide bridge was structured by optical lithography and ion-beam etching. Transport measurements were performed in the standard four-point configuration using a 9\,T physical property measurement system (PPMS) by \textit{Quantum Design}.

\vspace{-5pt}\section{Monopole-monopole model}\vspace{-5pt}

\begin{figure}[t]
\includegraphics[width=\columnwidth]{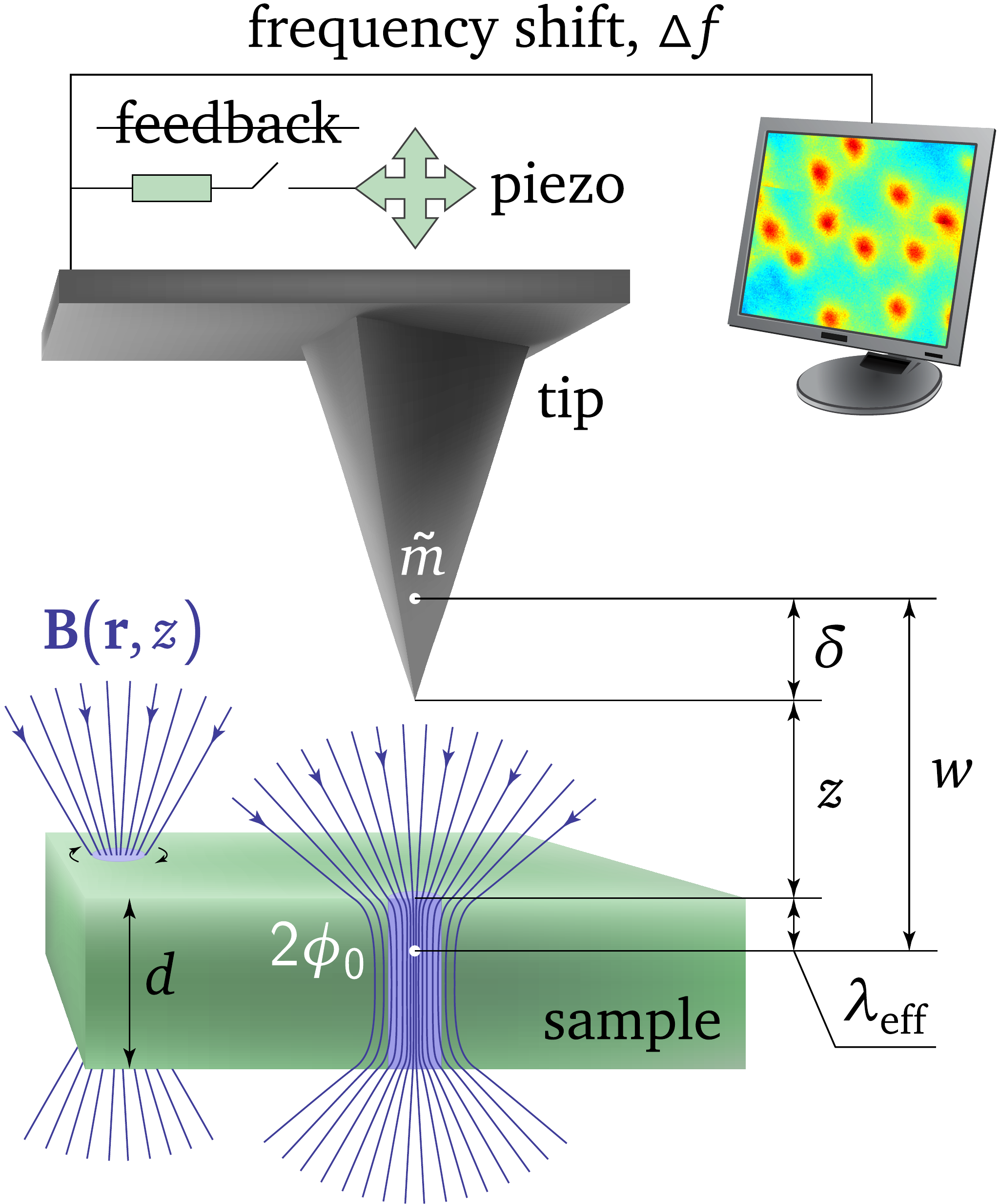}
\caption{A schematic of the MFM imaging procedure and an illustration of the monopole-monopole model. The magnetized MFM tip, driven by a piezo element, scans above the surface of the sample at a given distance $z$. During measurement, the feedback loop, which is typically used to stabilize the resonance frequency of the tip during topographic imaging, is deactivated. One measures a shift of the resonance frequency, ${\scriptstyle\Delta}f\!$, induced by the magnetic field of the vortices, $\mathbf{B}(\mathbf{r},z)$. In the monopole-monopole model, described in the text, both the tip and the vortex are approximated by magnetic monopoles at distances $z+\delta$ and $\lambda_{\rm eff}$ from the surface of the sample, respectively.\vspace{-0.5em}}
\label{Fig:Figure2}
\end{figure}

The magnetic moment of the MFM tip (see sketch in Fig.~\ref{Fig:Figure2}) can be well approximated to the first order by a magnetic monopole characterized by a ``magnetic charge'' $\tilde{m}$ located at a distance $\delta$ from the sharp end of the tip pyramid~\cite{Hug98}. The field distribution from a single flux quantum, $\phi_0=2.07\times10^{-15}$\,T\,m$^2$, measured just above the surface of the superconductor, is also similar to the magnetic field emanated by a magnetic monopole of $2\phi_0$, located at the depth $\lambda_{\rm eff}=1.27\lambda$ below the surface~\cite{Car00}, with $\lambda$ being the magnetic penetration depth. Hence, the magnetic induction of the vortex, $\textbf{B}(\mathbf{r},z)$, taken at a distance $z$ above the surface, can be approximated by
\begin{equation}
\textbf{B}(\mathbf{r},z) = \frac{\phi_0}{2\pi}\frac{(\textbf{r}-\textbf{r}_0)+(z+1.27\lambda)\,\textbf{e}_z}{\big((\mathbf{r}-\mathbf{r}_0)^2+(z+1.27\lambda)^2 \big)^{3/2}},
\end{equation}
where $\textbf{e}_z$ is the unit vector orthogonal to the film, $\mathbf{r}_0$ is the position of the vortex core and $\mathbf{r}$ is the radial distance from its center. This leads us to the tip-vortex interaction force $\textbf{F}(\mathbf{r},z)$ in the monopole-monopole model,
\begin{equation}
\textbf{F}(\mathbf{r},z)=\tilde{m}\textbf{B}(\mathbf{r},z).
\end{equation}
Taking into account that the shift of the resonance frequency of the cantilever measured by MFM, ${\scriptstyle\Delta}f\!$, is proportional to the $z$-derivative of the normal component of the force that acts between the tip and the sample \cite{Hug98, Alb91}, $\partial_z F_z(\mathbf{r},z)$, one obtains the following expression for the measured signal:
\begin{equation}
{\scriptstyle\Delta}f = -\frac{f_0}{2k} \frac{\tilde{m}\phi_0}{2\pi} \frac{(\mathbf{r}-\mathbf{r}_0)^2-2(z+1.27\lambda+\delta)^2}{((\mathbf{r}-\mathbf{r}_0)^2+(z+1.27\lambda+\delta)^2)^{5/2}}.
\label{Eq:ResShift}
\end{equation}
\noindent As the magnetic induction of the vortex is maximal at the center and decays rapidly with $r$, the strongest interaction in $z$ direction between the tip and the vortex occurs when the tip passes the center of the vortex. Thus, the maximal $z$-component of this force is reached at $r=r_0$:
\begin{equation}
\text{max}(F_z)_{r=r_0} = \frac{\tilde{m}\phi_0}{2\pi} \frac{1}{(z+1.27\lambda+\delta)^2}
\label{Eq:Fmax}
\end{equation}
The benefit of the monopole-monopole model is that all spatial parameters of the problem ($z$, $\lambda_{\rm eff}$ and $\delta$) enter Eqs.\,\ref{Eq:ResShift} and \ref{Eq:Fmax} additively, hence the sum in the denominator can be redefined as an effective tip-sample distance $w = z + \lambda_{\rm eff} + \delta$. It represents the distance between imaginary magnetic monopoles within the tip and the vortex, as illustrated by white dots in Fig.\,\ref{Fig:Figure2}.

The ratio between the lateral and vertical forces that act during scanning typically varies from 0.3 for a tip with a less sharp pyramid~\cite{Wad92} to $2/3\sqrt{3}\approx0.38$ for the monopole-monopole model~\cite{Str08, Wad92}. This results in the following maximum lateral component of the tip-vortex interaction force:
\begin{equation}
\text{max}(F_\text{lat}) \approx 0.38 \cdot \text{max}(F_z).
\label{Eq:Flat}
\end{equation}
Obviously, non-invasive imaging of vortices by MFM is possible only as long as the vortices are pinned. The tip-vortex interaction force can be accurately tuned during scanning by varying the tip-sample separation $z$~\cite{Str08}. If this force exceeds the pinning force of an individual vortex at a natural or artificial defect, the vortex can be dragged away from its initial position. Likewise, if the tip-sample distance is kept constant, an increase in temperature can lead to the local depinning of individual flux lines during scanning due to the temperature dependence of the pinning force.

\vspace{-5pt}\section{Local depinning of individual flux lines}\vspace{-5pt}

\begin{figure}
	\centering
		\includegraphics[width=\columnwidth]{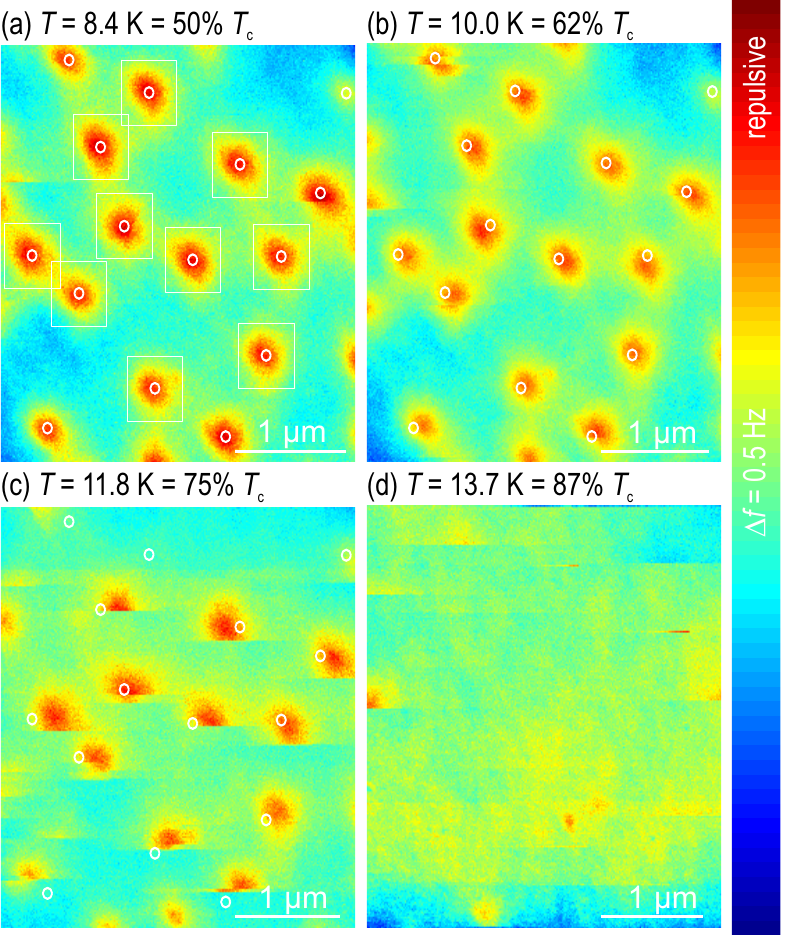}
	\caption{Vortex images measured at a distance $z=30$\,nm from the surface of a NbN film after field-cooling in $-3$\,mT at various temperatures: (a) 50$\%\,T_{\rm c}$, (b) 62$\%\,T_{\rm c}$, (c) 75$\%\,T_{\rm c}$, and (d) 87$\%\,T_{\rm c}$. The slow scanning direction is top to bottom. Depinning of vortices by the magnetic tip can be seen in panels (c) and (d). \vspace{-0.5em}}
	\label{Fig:Figure3}
\end{figure}

We have imaged the temperature dependence of the vortex distribution after field-cooling the sample to $\sim$\,8\,K in a vertical applied magnetic field of $\mu_0 H_z=-3$\,mT (Fig.\,\ref{Fig:Figure3}), and subsequently increasing the temperature in steps of 2\,K. The tip-sample distance was kept constant at $z=30$\,nm, so that Eq.\,(\ref{Eq:Fmax}) can be applied. The negative direction of the field corresponds to the magnetic repulsion between the MFM tip and the vortex, and therefore the vortices appear as red (dark-grey) objects in the false-color images presented in Fig.\,\ref{Fig:Figure3}. In all panels, the white circles depict the positions of the vortex cores at base temperature, to emphasize the tip-induced changes in their position as the temperature is increased. These positions were determined by a two-dimensional (2D) fitting procedure \cite{Ino10} applied to the regions indicated by rectangles. The benefit of this method is that, in principle, it allows for a subpixel resolution of the fitting, given that the noise in the measured data is sufficiently low.

At temperatures not exceeding $\sim50\%\,T_{\rm c}$ [Fig.\,\ref{Fig:Figure3}\,(a)], noninvasive imaging of vortices takes place, indicating that the pinning exceeds the lateral thrust of the MFM tip. Statistical image analysis, such as described by Inosov \textit{et al.}~\cite{Ino10}, reveals that the vortices form a highly disordered hexagonal lattice due to the pinning by natural defects.

At higher temperatures, the decreasing contrast of the vortex profile signifies a natural increase in the penetration depth, $\lambda$, that characterizes the decay of the magnetic field outside of the vortex core. At 62\%\,$T_{\rm c}$ [Fig.\,\ref{Fig:Figure3}\,(b)], most vortices are still in their original positions, implying that the tip-vortex interaction force is still lower than the typical pinning force of a single vortex. Only 2 out of 14 vortices, visualized in the figure, have been irreversibly dragged away from their initial positions (white circles) to the nearest pinning sites with higher pinning potentials. This indicates the existence of a slightly modulated pinning landscape in the NbN film and, hence, a spatial variation of the pinning force, as expected for natural defects.

\begin{table}[b]\normalsize
\begin{center}
    \begin{tabular}{l@{~}l@{~}|@{~~}l@{~~}l@{~~}l@{~~}l@{~~}l}\toprule
    \multicolumn{2}{l}{Temperature\,(K):} & ~~~8.4  &  ~~~9.1 & ~~10.0  & ~~11.8\\
    \midrule
	$w$                                              & (nm) & 305(11)  & 308(10) & 316(10)  & 383(23)\!\!\! \\
    ${\scriptstyle\Delta}f(\mathbf{r}=\mathbf{r}_0)$ & (Hz) & 0.182(4) &         & 0.134(3) & 0.138(6)\!\!\! \\
	$\mathrm{max}(F_{\rm lat})$                      & (pN) & 0.74(3)  &         & 0.56(2)  & 0.70(5) \\
	\bottomrule
\end{tabular}
\end{center}\vspace{-6pt}
\caption{Temperature dependence of the effective tip-sample distance in the monopole-monopole model, $w=z+1.27\lambda+\delta$; the fitted peak amplitude from Fig.\,\ref{Fig:Figure4}, ${\scriptstyle\Delta}f(\mathbf{r}=\mathbf{r}_0)$; and the corresponding lateral tip-vortex interaction, $\mathrm{max}(F_{\rm lat})$.}
\label{Tab:Table1}
\vspace{-11.5pt}
\end{table}

\begin{figure}
	\centering
		\includegraphics[width=\columnwidth]{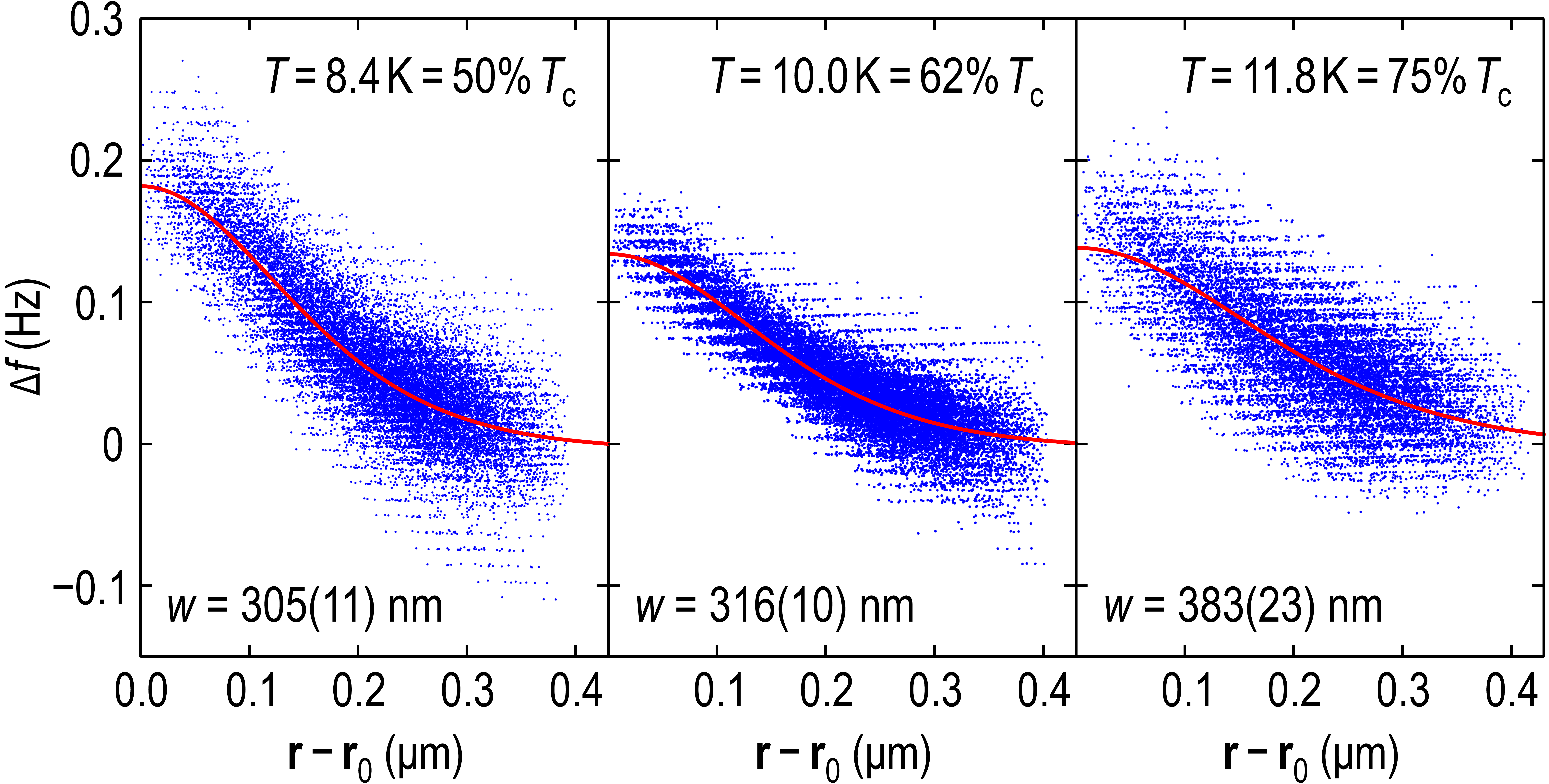}
	\caption{Temperature dependence of the averaged vortex profile. Blue points are the measured signal, solid lines are least-squares fits to the monopole-monopole model.\vspace{-0.5em}}
	\label{Fig:Figure4}
\end{figure}

At 75\%\,$T_{\rm c}$ [Fig.\,\ref{Fig:Figure3}\,(c)], the movement of nearly every vortex by the MFM tip can be seen. Indeed, because most vortices are irreversibly dragged away by the tip as it passes close to the core, such vortices appear half-cut in the image. Consequently, at this temperature the pinning force for the majority of the vortices is equal to the maximal lateral force exerted by the MFM tip onto the vortex, Eq.\,(\ref{Eq:Flat}). Only one vortex at the bottom-right part of the image remains stable, evidencing the locally enhanced pinning force at this position. On the other hand, three other vortices are fully dragged away as soon as the tip starts crossing their field lines. At even higher temperatures [Fig.\,\ref{Fig:Figure3}\,(d)], vortices can no longer be imaged. Here the vortices are being continuously dragged by the tip during scanning.

For a quantitative analysis of the vortex profiles, their core positions $\mathbf{r}_0$ were first determined using 2D fitting. To gather sufficient statistics for the application of the monopole-monopole model, the signal ${\scriptstyle\Delta}f$ from within the neighborhood of every vortex (white rectangles in Fig.\,\ref{Fig:Figure3}) has been plotted vs.\,$|\mathbf{r}-\mathbf{r}_0|$, as shown in Fig.\,\ref{Fig:Figure4}. In this figure, every data point corresponds to a pixel in the original MFM image, whereas data points originating from different vortices are combined in one plot. The resulting clusters of points can be fitted to Eq.\,(\ref{Eq:ResShift}) (solid lines) to obtain the average value of $w=z+1.27\lambda+\delta$ for every temperature with a sufficiently small statistical error. The results of these fits are summarized in Table\,\ref{Tab:Table1}.

\vspace{-5pt}\section{Local pinning forces and the penetration depth}\vspace{-5pt}

\begin{figure}[t]\vspace{-0.5em}
\includegraphics[width=\columnwidth]{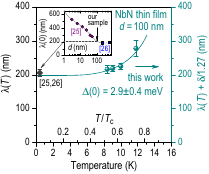}
\caption{Temperature dependence of $\lambda(T)+\delta/1.27$ values (plotted on the right scale) in comparison to $\lambda(0)$ (left scale). The solid line is a fit to an empirical model with a single $s$-wave gap \cite{GapModel}. The value of $\lambda(0)$ for our film thickness was obtained by an interpolation of the $d$-dependent literature data \cite{Kamlapure10, Wan96}, as shown in the inset. The solid grey line is an empirical fit. A good agreement of our results with the published value indicates that the $\delta/1.27$ correction is negligible in our case.\vspace{-0.8em}}
\label{Fig:Figure5}
\end{figure}

Now we can proceed to the quantitative estimation of the pinning force. Combining Eq.\,(\ref{Eq:ResShift}) through (\ref{Eq:Flat}), we obtain
\begin{equation}
\text{max}(F_\text{lat}) \approx 0.38\,\frac{kw}{f_0}\cdot{\scriptstyle\Delta}f(\mathbf{r}=\mathbf{r}_0).
\end{equation}
Substituting the fitting results from Fig.\,\ref{Fig:Figure4} and the known parameters of the tip into this expression, one can calculate the temperature-dependent lateral tip-vortex forces that are listed in Table\,\ref{Tab:Table1}, with an average value of $0.67\pm0.09$\,pN \cite{Comment}. As we already know that the local pinning force decreases sufficiently to allow depinning of most vortices at $T\approx12\,\text{K}=75\%\,T_{\rm c}$, we can use the calculated value as an estimate of the mean local pinning force at this temperature,\vspace{-0.2em}
\begin{equation}
F_\text{p}\,(T=12\,\text{K}) \approx (0.67\pm0.09)\,\text{pN}.\vspace{-0.2em}
\end{equation}

The results of the same fit also provide a local probe for the temperature-dependent penetration depth \cite{Roseman01, Luan10, Kamlapure10, Wan96}, in our case given by $\lambda(T)\approx[w(T)-z-\delta]/1.27$. The first temperature-independent parameter $z\ll\lambda$ is fixed during measurement and known, so it can be easily subtracted (Fig.\,\ref{Fig:Figure5}). The second parameter, $\delta$, being a property of the tip and dependent both on temperature and the stray field of the sample, usually has a large uncertainty and requires a special calibration of the tip in order to be determined \cite{TipCalibration}. It may lead to a non-negligible constant offset of the measured penetration depth from its true value. To quantify the $\delta/1.27$ correction in the present work, we resorted to a comparison of our temperature-dependent data (shown by spheres in Fig.\,\ref{Fig:Figure5}) to the low-temperature value of $\lambda(0)\approx205$\,nm (dark-gray diamond in Fig.\,\ref{Fig:Figure5}). The latter has been obtained by interpolating the directly measured $\lambda(0)$ values from the literature for different film thicknesses \cite{Kamlapure10, Wan96}, as shown in the inset of Fig.\,\ref{Fig:Figure5}, and taking the intermediate value at $d=100$\,nm. Good agreement between this reference value of $\lambda(0)$ and our $\lambda(T)+\delta/1.27$ points indicates that the $\delta/1.27$ correction is negligible in our case.

\begin{figure}[b]
	\centering
		\includegraphics[width=\columnwidth]{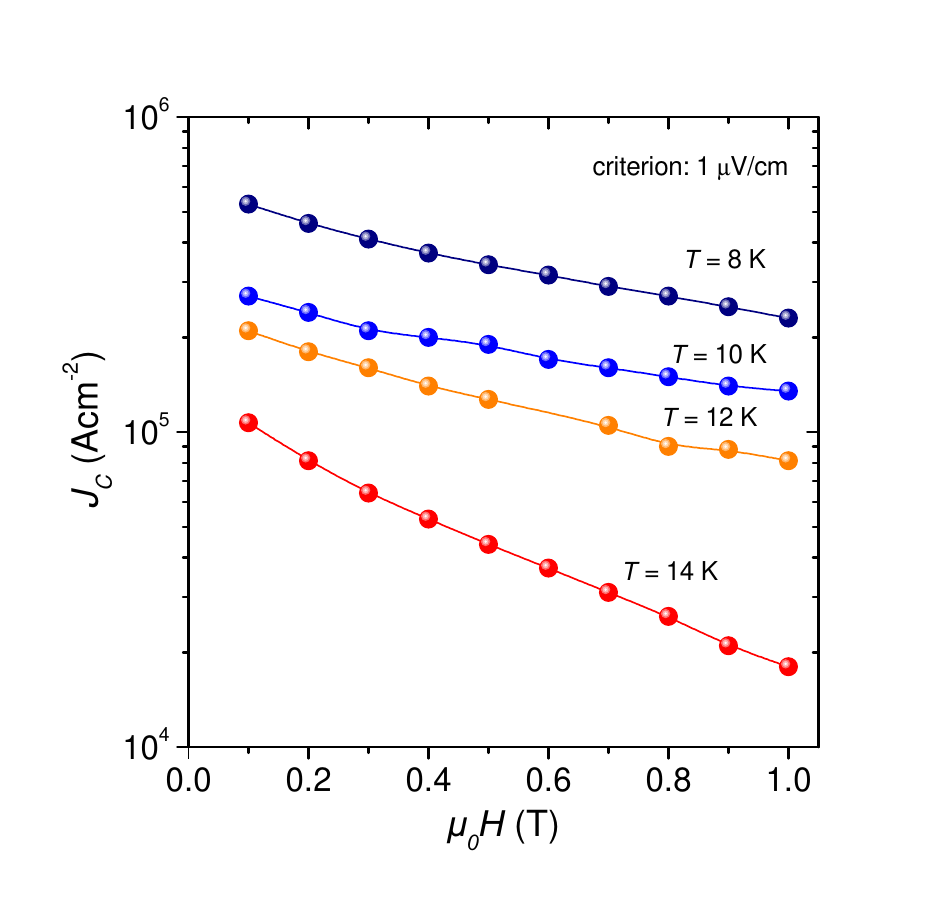}
	\caption{Dependence of the critical current density, $J_{\rm c}$, on applied magnetic field for different temperatures: 8\,K, 10\,K, 12\,K and 14\,K.\vspace{-1.2em}}
	\label{Fig:Figure6}
\end{figure}

The solid line in Fig.\,\ref{Fig:Figure5} is an empirical fit to the formula of Evtushinsky \textit{et al.} \cite{GapModel}, which gives an analytical relationship between $\lambda(T)$ and the SC gap $\Delta(0)$. For a conventional superconductor with a single isotropic $s$-wave gap, it becomes\vspace{-0.05em}
\begin{equation}
\lambda(T)=\lambda(0)\!\left[\!1-M\Biggl(\!\!\frac{\Delta(T)}{k_{\rm B}T}\!\!\Biggr)\!\right]^{-1/2}\hspace{-2em},
\label{Eq:PenetrationDepth}
\end{equation}
where $\lambda(0)$ depends only on the band structure, whereas all the temperature-dependent quasiparticle effects are included in the approximant function \cite{GapModel}
\begin{equation}
M(t)=4\,(\mathrm{e}^{t/2}+\mathrm{e}^{-t/2})^{-2}\sqrt{\piup\kern.5pt t\kern-1pt/\kern-.5pt 8+1/(1+\piup\kern.5pt t\kern-1pt/\kern-.5pt8)}.
\end{equation}
The temperature dependence of the SC gap in Eq.\,(\ref{Eq:PenetrationDepth}) is approximated by \cite{GrossChandrasekhar86}\vspace{-0.3em}
\begin{equation}
\Delta(T)=\Delta_0\,\tanh\biggl(\!\frac{\piup}{2}\sqrt{T_{\rm c}/T-1}\,\biggr).\vspace{-0.3em}
\label{Eq:TdepGap}
\end{equation}
The resulting fit yields a value of $\Delta(0)=(2.9\pm0.4)$\,meV = $(2.1 \pm 0.3)\,k_{\rm B}T_{\rm c}$, which perfectly agrees with direct tunneling measurements \cite{Wan96, Kamlapure10} and is slightly above the weak-coupling limit of $1.76\,k_{\rm B}T_{\rm c}$ predicted by the Bardeen-Cooper-Schrieffer (BCS) theory \cite{BCS57}.

\vspace{-5pt}\section{Global estimate of the pinning forces}\vspace{-5pt}

While MFM provides access to the pinning force of individual vortices, the global characterization methods, such as transport or magnetization measurements, explore the collective behavior of the flux-line lattice. They evaluate the mean pinning force within the whole sample volume, considering also the elastic interaction between individual flux lines as well as the collective pinning~\cite{Tin96}. For small magnetic fields, $B \ll \mu_0H_{c2}$, the distance between vortices is larger than $\lambda$. Such vortices can be treated as independent non-interacting objects. In this limit, collective effects can be ignored and the pinning force per vortex can be calculated as $F^{\rm g}_{\rm p}/N$, where $F^{\rm g}_{\rm p}$ is the average pinning force and $N$ is the number of vortices within the sample surface.

\begin{table}[t]\vspace{1ex}
\begin{tabular}{l@{~}l@{~~}|@{~~}c@{~~}c@{~~}c@{~~}c}
    \toprule
    \multicolumn{2}{l}{Temperature\,(K):} & 8 & 10 & 12 & 14 \\
    \midrule
    $J_{\rm c}$     & (10$^{5}$A/cm$^2$)  & 10  & 5  & 3  &  2   \vspace{2pt}\\
    $F^{\rm g}_{\rm p}/N$ & (pN)               & 2.07 & 1.04& 0.62& 0.42 \\
    \bottomrule
\end{tabular}
\caption{Critical current density and average pinning force per vortex at different temperatures evaluated from the transport data in Fig.\,\ref{Fig:Figure6} for $B=0$.\vspace{-1.3em}}
\label{Tab:Table2}
\end{table}

The pinning force is equal to the maximal sustainable Lorentz force that does not move vortices while the current flows~\cite{Tin96}:\vspace{-1.0em}
\begin{equation}
F^{\rm g}_{\rm p}(B)= V J_{\rm c}\,B = S d J_{\rm c}\,B,
\end{equation}
\noindent where $J_{\rm c}$ is the critical current density, $S$ is the surface area of the sample and $d$ is the sample thickness. The number of vortices is $N = BS/{\phi_0}$, hence the pinning force per vortex is
\begin{equation}
F^{\rm g}_{\rm p}/N  = J_{\rm c}\,d \phi_0.
\end{equation}

Dependence of $J_{\rm c}$ on the applied magnetic field for temperatures between 8 and 14\,K is presented in Fig.\,\ref{Fig:Figure6}. The resulting temperature dependence of the pinning force per vortex $F^{\rm g}_{\rm p}/N$ calculated from these $J_{\rm c}(H)$ curves is given in Table~\ref{Tab:Table2}. One can immediately appreciate the agreement between the value of $F^{\rm g}_{\rm p}/N=0.62$\,pN that resulted from the transport measurements at $T=12$\,K with that of $(0.67\pm0.09)$\,pN that we extracted earlier from the MFM data at a similar temperature. Taking into account that the pinning force varies by nearly a factor of 5 in the studied temperature range, such an agreement within 8\% between local and global measurements is indeed remarkable.

\vspace{-5pt}\section{Summary}\vspace{-5pt}

To conclude, we found perfect agreement between the values of the pinning force per vortex, estimated from local depinning of individual vortices by the MFM tip and globally from the critical current measurements. We demonstrated that for low fields, $B \ll \mu_0 H_{\rm c2}$, MFM is a powerful and reliable method to probe the local space variation of the pinning landscape. The monopole-monopole model, originally derived for $d > 4\lambda$~\cite{Car00}, proved to be successful even for thin films with a thickness comparable to the penetration depth. With this knowledge, the quality of such very thin films, that are actually employed for the application in SC bolometers \cite{Sem05}, can be perfectly analyzed using magnetic force microscopy.

Finally, we used accurate 2D fitting of the vortex profiles to extract the London penetration depth of the NbN film and the SC energy gap. The statistical errors of this method are small enough to ensure that the extracted gap $\Delta(0)=(2.9\pm0.4)$\,meV agrees with the directly measured values \cite{Wan96, Kamlapure10}. Although similar methods of extracting the gap amplitude from the muon-spin rotation ($\mu$SR), small-angle neutron scattering (SANS), microwave surface-impedance (MSI), or magnetization measurements of the temperature-dependent penetration depth are well developed \cite{InosovReview}, their application to the analysis of temperature-dependent LT-MFM images is only becoming a standard practice \cite{Luan10}.

\vspace{-5pt}\section*{Acknowledgments}\vspace{-5pt}
The authors acknowledge discussions with E.~H. Brand and are grateful to K.~Osse, S.~Sieber and I.~Mönch for sample preparation, to D.~Meier and M.~Gründlich for technical assistance in sample polishing, to U.~Wolff for MFM tip preparation, and to Kawia Scharle for the design of Fig.\,2.


\begin{thebibliography}{30}
\bibitem{Comment} In our temperature range, the changes of $\text{max}(F_{\rm lat})$ due to the temperature variations of $\lambda$ in Eq.\,(\ref{Eq:Fmax}) are smaller than the statistical uncertainties of the measured values given in Table\,\ref{Tab:Table1}. Therefore, such averaging is justified. In a broader temperature range, fitting the measured values to Eq.\,(\ref{Eq:Fmax}) might be necessary.
\end{thebibliography}
\end{document}